# Discovery of the Zintl-phosphide BaCd$_2$P$_2$ as a long carrier lifetime and stable solar absorber

Zhenkun Yuan,[1] Diana Dahliah,[2,3] Muhammad Rubaiat Hasan,[4] Gideon Kassa,[1] Andrew Pike,[1] Shaham Quadir,[5] Romain Claes,[2] Cierra Chandler,[6] Yihuang Xiong,[1] Victoria Kyveryga,[4,7] Philip Yox,[4,7] Gian-Marco Rignanese,[2] Ismaila Dabo,[6] Andriy Zakutayev,[5] David P. Fenning,[8] Obadiah G. Reid,[5,9] Sage Bauers,[5] Jifeng Liu,[1] Kirill Kovnir,[4,7] and Geoffroy Hautier[1,*]

*Corresponding author. Email: geoffroy.hautier@dartmouth.edu

1 Thayer School of Engineering, Dartmouth College, Hanover, NH 03755, USA
2 Institute of Condensed Matter and Nanosciences, Université Catholique de Louvain, Chemin des Étoiles 8, B-1348, Louvain-la-Neuve, Belgium
3 Department of Physics, An-Najah National University, Nablus, Palestine
4 Department of Chemistry, Iowa State University, Ames, IA 50011, USA
5 National Renewable Energy Laboratory, Golden, CO 80401, USA
6 Department of Materials Science and Engineering, and Materials Research Institute, The Pennsylvania State University, University Park, PA 16802, USA
7 Ames National Laboratory, U.S. Department of Energy, Ames, IA 50011, USA
8 Department of Nanoengineering, UC San Diego, La Jolla, CA 92093, USA
9 University of Colorado Boulder, Boulder, CO 80309, USA

## Abstract

Thin-film photovoltaics offers a path to significantly decarbonize our energy production. Unfortunately, current materials commercialized or under development as thin-film solar cell absorbers are far from optimal as they show either low power conversion efficiency or issues with earth-abundance and stability. Entirely new and disruptive materials platforms are rarely discovered as the search for new solar absorbers is traditionally slow and serendipitous. Here, we use first principles high-throughput screening to accelerate this process. We identify new solar absorbers among known inorganic compounds using considerations on band gap, carrier transport, optical absorption but also on intrinsic defects which can strongly limit the carrier lifetime and ultimately the solar cell efficiency. Screening about 40,000 materials, we discover the Zintl-phosphide BaCd$_2$P$_2$ as a potential high-efficiency solar absorber. Follow-up experimental work confirms the predicted promises of BaCd$_2$P$_2$ highlighting an optimal band gap for visible absorption, bright photoluminescence, and long carrier lifetime of up to 30 ns even for unoptimized powder samples. Importantly, BaCd$_2$P$_2$ does not contain any critical elements and is highly stable in air and water. Our work opens an avenue for a new family of stable, earth-abundant, high-performance Zintl-based solar absorbers. It also demonstrates how recent advances in first principles computation can accelerate the search of photovoltaic materials by combining high-throughput screening with experiment.

## Introduction

Photovoltaics (PV) will play a central role in the ongoing energy transition towards a sustainable, low-carbon-emission society.[1-3] In that regard, thin-film solar cells have remarkable prospects, because their manufacturing can require smaller upfront capital investment, less energy input and produces fewer greenhouse gas emissions than today's dominant silicon-based technology. Thin-film technologies can also offer compatibility with flexible substrates and band-gap tuning opportunities leading to building-integrated power generation or tandem devices in combination with silicon PV. Among the different materials used as solar cell absorbers, CdTe and Cu(In,Ga)Se$_2$ (CIGS) have achieved high power conversion efficiency of over 22% and have been commercialized for almost two decades.[4,5] Unfortunately, CdTe and CIGS rely on critical elements, such as tellurium and indium, raising concerns for multi-terawatt scale deployment.[6,7] Over the last decade, earth-abundant thin-film materials such as Cu$_2$ZnSn(S,Se)$_4$ (CZTS) and, in particular, lead halide perovskites (e.g., MAPbI$_3$, where MA=CH$_3$NH$_3^+$) have emerged. However, CZTS lags behind commercial thin-film technologies in efficiency, while halide perovskites suffer from long-term stability issues.[8-14]

Large-scale deployment of thin-film PV demands high-efficiency, stable, and low-cost solar absorbers that are yet to be discovered. For over half a century, only a limited number of materials have been explored for thin-film PV applications,[15-19] as conventional materials discovery is an intrinsically slow process. Most of the materials research efforts in the thin-film PV field have been devoted to modifying known semiconductors (cation substitution in CZTS,[20-23] mixed-cation/mixed-halide perovskites,[24-26] …) but the recent advent of halide perovskites demonstrates that an unexpected new family of PV materials can still be found and revolutionize the field. This motivates further "out-of-the-box" explorative work, especially towards more stable materials than the halide perovskites.

High-throughput (HT) computational screening can significantly accelerate materials discovery and is emerging in the PV field.[27-37] While earlier HT searches for solar absorbers focused on bulk properties, such as band gap, effective mass, and optical absorption coefficient,[27-35] more complex properties related to point defects have started to be integrated into the screening as well.[36,37] Considering defect-assisted nonradiative carrier recombination is essential in the search for new high-performance solar absorbers. Indeed, designing and discovering materials with "defect tolerance" and long carrier lifetime has become an important target for the PV community.[38-45] Yet, only a few HT studies have included defect properties in the screening process, and their focus so far is mainly on copper-based inorganic materials[36] and perovskite compounds.[37] Moreover, these computational predictions have not yet lead to experimental realization of a proposed solar absorber.

Here, we identify the BaCd$_2$P$_2$ as a long carrier lifetime and stable solar absorber using HT computational screening followed by bulk synthesis and experimental characterization. We perform HT search to screen about 40,000 inorganic materials obtained from the Materials Project database,[46] most of which are registered in the Inorganic Crystal Structure Database (ICSD).[47] We find a handful of materials combining a suitable band gap, small effective masses, and promising defect properties which will not cause strong nonradiative carrier recombination. Among these promising candidates, we select the Zintl-phosphide $BaCd_2P_2$ and explicitly show that the computed nonradiative recombination rates in $BaCd_2P_2$ are better than or comparable to those in high-efficiency solar absorbers such as the halide perovskites. We experimentally synthesize and characterize $BaCd_2P_2$, showing that this material is highly stable in air and water. Bright photoluminescence (PL) and time-resolved microwave conductivity (TRMC) measurements confirm the computed direct band gap of 1.45 eV and present a promising long carrier lifetime of up to 30 ns. All of these results indicate that $BaCd_2P_2$ is a promising high-performance solar cell absorber with the potential to open a new avenue in PV for an entire family of Zintl $AM_2X_2$ solar absorbers (where A and M are +2 ions and X is a pnictogen).

## Results and Discussion

### High-throughput computational screening for high-efficiency solar absorbers

Our starting point is the Materials Project electronic structure database. Without constraints on any specific chemistry, we obtain 39,659 materials with full data on crystal structure, electronic band structure, and effective masses. We adopt a tiered computational screening approach (very similar to our previous work[36]) to search for solar cell absorbers among these materials (see Note S1 of the supplemental information). Including electronic and defect properties in HT screening requires combining calculations at different levels of theory from (semilocal) Density Functional Theory (DFT) to more accurate but computationally expensive Heyd-Scuseria-Ernzerhof (HSE) hybrid functional.[48] Our screening consists of considering first DFT band gaps and effective masses, followed by a refinement of the band gap using the HSE hybrid functional and then defect computations combining DFT and HSE. Finally, we perform full HSE defect calculations for 19 candidates, and predict their solar-cell efficiency using an extended detailed-balance (i.e., Shockley-Queisser) model which accounts for defect-assisted nonradiative carrier recombination.[36,49,50] At this screening stage we examine only vacancies and cation-cation antisites, without considering interstitials and cation-anion antisites. For nonradiative recombination, we assume a constant carrier capture cross section for all defects. This effectively penalizes materials with low-formation-energy, deep defects. Previous work has shown that this model can discriminate between high-performance solar absorbers and materials that are unlikely to realize high efficiency because of their defect properties.[36,49,50]

The 19 candidates together with 12 established solar absorbers are displayed in Figure 1 with their theoretical power conversion efficiency and estimated material cost (see Note S1 for more details). Additional metrics beyond cost related to production and earth abundance can also be found in Figures S2. Given that stability issues are hampering the deployment of halide perovskite solar cells, we have taken into account stability in selecting candidate materials. Unfortunately, there is no easy-to-compute criterion for assessing stability in realistic conditions. We can estimate the thermodynamic stability (energy above hull) of a material versus competing phases from DFT, but this is a poor indicator of stability in air and with respect to moisture. In fact, much of non-oxides (e.g., phosphides) show thermodynamic instability in contact with oxygen. Kinetic effects and surface passivation are likely to be key factors for their stability but are currently poorly assessed by theory. To assess stability, we rely on qualitative estimates based on the solid-state-chemistry experience of the authors and by following the literature report on these materials including the work reporting their synthesis. For instance, for the many alkali and alkaline-earth metal pnictides (phosphides, antimonides, …) in the list, we have used a simple empirical rule: the higher the alkali or alkaline-earth per formula unit, the less stable the material will be. For example, according to this rule, KZnP (33%) should be less stable than $BaCd_2P_2$ (20%). Our stability assessment is conveyed in Figure 1 through the color of the data point.

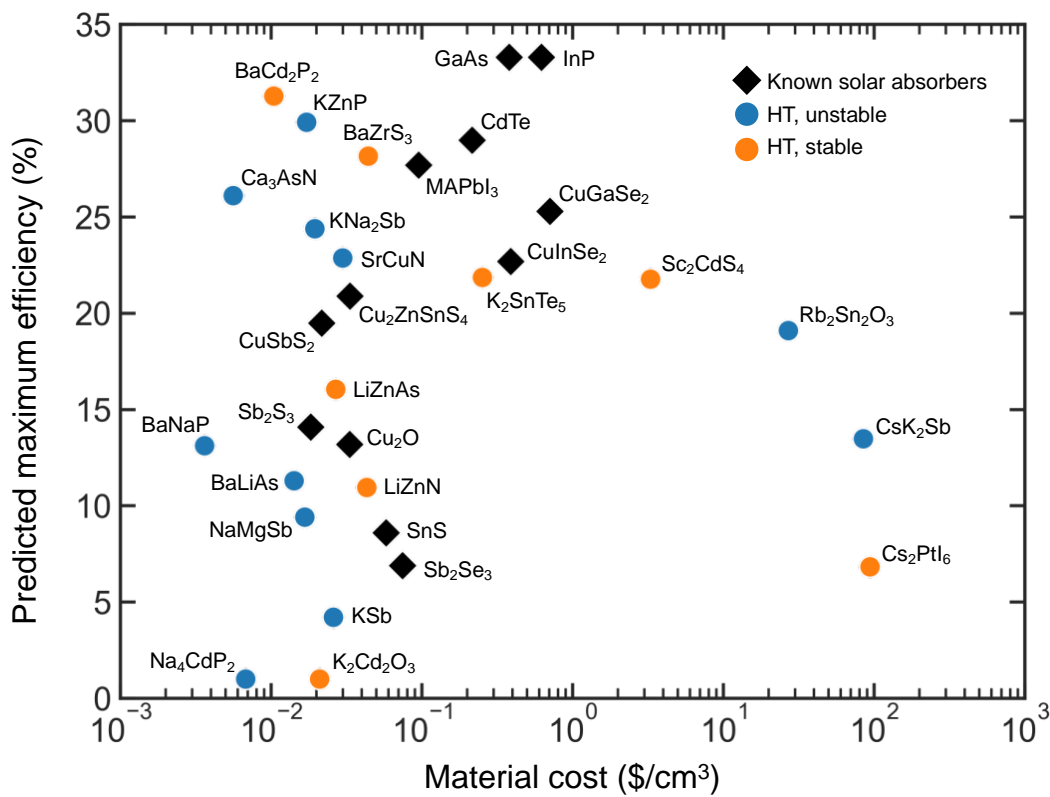

**Figure 1 Theoretical power conversion efficiency versus material cost for 19 potential thin-film solar cell absorbers identified from high-throughput (HT) screening** Black diamonds represent known thin-film solar absorbers for comparison purposes. Blue circles represent materials that are likely to be sensitive to air and moisture. Orange circles represent materials that are probably stable in air and moisture.

The most promising candidates when combining power conversion efficiency, materials cost, and stability are BaCd$_2$P$_2$ and BaZrS$_3$. The BaZrS$_3$ perovskite has recently been studied for thin-film PV applications.[51,52] In contrast, BaCd$_2$P$_2$ is classified as a Zintl compound and has received little attention in the PV field since it was experimentally reported in the 1980s.[53,54] While BaCd$_2$P$_2$ has recently been shown to possess a suitable band gap and small effective masses,[32,55] our HT screening, which includes defect-assisted nonradiative recombination, makes it stand out from the vast number of materials with suitable band gaps. Other candidates such as the spinel ScCd$_2$S$_4$ or K$_2$SnTe$_5$ have cost and abundance issues. They are still of scientific interest and substitutional strategies could be considered to reduce or even eliminate the use of the critical elements (Sc and Te). Other interesting candidates such as KZnP, Na$_2$KSb, SrCuN, and Ca$_3$AsN are excluded due to stability concerns, confirmed by their experimental synthesis reports.[56-59] Based on these considerations, we select BaCd$_2$P$_2$ as the most promising solar absorber candidate. In the following, we provide a more detailed study of the optoelectronic and defect properties of BaCd$_2$P$_2$.

## BaCd$_2$P$_2$ computed bulk optoelectronic properties

BaCd$_2$P$_2$ crystallizes in the $P\bar{3}m1$ CaAl$_2$Si$_2$ Zintl-type structure[60] (Figure 2A). The crystal structure consists of alternating layers of tetrahedrally coordinated Cd and octahedrally coordinated Ba. This structure type has attracted substantial interest for thermoelectric applications, e.g., the antimonides Mg$_3$Sb$_2$ and CaMg$_2$Sb$_2$.[61-65] Figure 2B shows the BaCd$_2$P$_2$ band structure calculated with HSE. The direct band gap value of 1.45 eV is ideal for single-junction solar cells according to the detailed-balance model. There is a competing indirect gap which should not be detrimental to PV performance and, instead, has been suggested to be beneficial to carrier lifetime.[66,67] The band structure exhibits dispersive valence and conduction bands. As a result, the effective masses are low for both electrons (0.11–0.74 $m_0$) and holes (0.44–0.63 $m_0$).[68] Further carrier mobility calculations including phonon scattering give a room-temperature mobility of 111–916 cm$^2$/Vs for electrons and 123–242 cm$^2$/Vs for holes depending on the crystallographic direction (Figure S3). The computed mobility is for a perfect single-crystal of BaCd$_2$P$_2$, and provides an upper bound for the mobilities in a polycrystalline film with impurities. While lower than in III-V semiconductors such as GaAs, the mobilities are comparable to those in many other solar absorbers such as CdTe,[69,70] CZTS,[71] and MAPI$_3$.[72] Besides, as shown in Figure 2C, the computed optical absorption coefficient BaCd$_2$P$_2$ is over 10$^4$ cm$^{-1}$ in the visible light range and is on par with those of known thin-film solar absorbers.

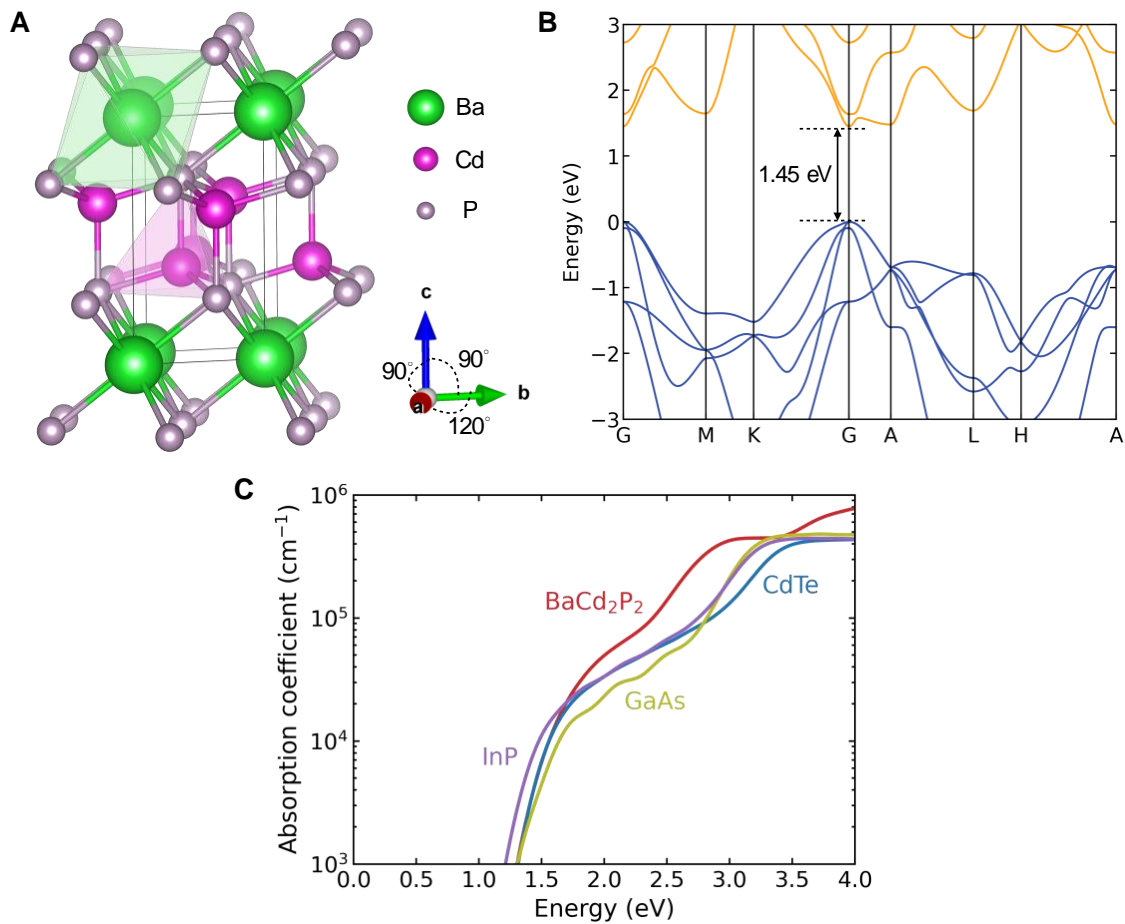

**Figure 2 Crystal structure, electronic band structure, and optical absorption of BaCd$_2$P$_2$** (A) Crystal structure of BaCd$_2$P$_2$. Green octahedra and pink tetrahedra show the phosphorus coordination of Ba and Cd, respectively. (B) HSE electronic band structure of BaCd$_2$P$_2$ (C) Calculated optical absorption spectra of BaCd$_2$P$_2$. The absorption calculation is at the DFT-GGA-PBE level, with the band gap corrected using the HSE value. Also shown are the calculated optical absorption spectra for GaAs, CdTe, and InP at the same level of theory.

## BaCd$_2$P$_2$ defect charge transition levels and dopability

Our HT screening has identified BaCd$_2$P$_2$ to be of interest not only for its band gap, carrier transport, and optical absorption but also for the absence of low-formation-energy, deep defects that can act as strong nonradiative recombination centers. In the initial screening, we considered only a limited types of intrinsic point defects. In the following, we provide a comprehensive first-principles study of intrinsic point defects in BaCd$_2$P$_2$. The electrical behavior of the defects is assessed based on their calculated formation energies and charge transition levels. The defect formation energy depends on the elemental chemical potentials which are related to synthesis conditions. The allowed chemical potentials of Ba, Cd, and P satisfy the stability condition of BaCd$_2$P$_2$, and are bound by the formation of different secondary phases in the Ba-Cd-P system (see Figure S4). Figures 3A and 3B plot the defect formation energies as a function of Fermi level under P-poor and P-rich conditions (Figure S4), respectively. Formation energies at other chemical-potential conditions can be found in Figure S5.

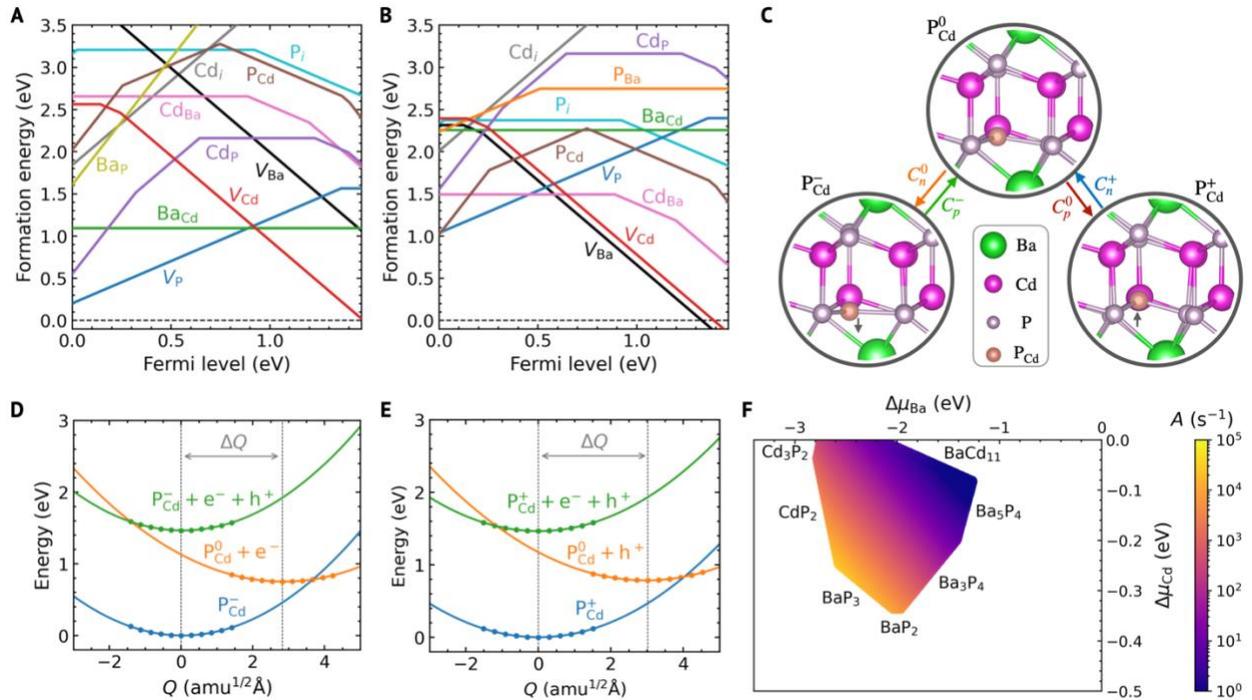

**Figure 3 Formation energies of intrinsic point defects, carrier capture at the P$_{Cd}$ antisite, and nonradiative carrier recombination in BaCd$_2$P$_2$** (A) and (B) HSE-calculated defect formation energies as a function of Fermi level under P-poor and P-rich conditions, respectively. (C) Local atomic structures of the P$_{Cd}$ antisite in the 0, −1, and +1 charge states (see also Figure S8D). The gray arrows indicate the relaxation of the phosphorus antisite when charged. Transition between two adjacent charge states involves two carrier capture processes, as indicated by the colored arrows and the corresponding capture coefficients; for the capture coefficients, the superscript denotes the initial charge state and the subscript denotes the carrier type ($n$, electron; $p$, hole). (D) and (E) Configuration coordinate diagrams describing the P$_{Cd}$ (0/−) and (0/+) charge-state transitions (see also Figures S8A–S8B), respectively. (F) Nonradiative recombination rates (denoted as $A$, in unit of s$^{-1}$) in intrinsic BaCd$_2$P$_2$ as a function of elemental chemical potentials in the ($\Delta\mu_{Ba}$, $\Delta\mu_{Cd}$) plane. The $A$ rates are given by $A = C_{tot} \times N_d$. Here, only the dominant nonradiative center P$_{Cd}$ is considered. The defect density ($N_d$) of P$_{Cd}$ is computed at 1000 K, while the total capture coefficient ($C_{tot}$) of P$_{Cd}$ is determined at 300 K.

Figures 3A and 3B show that vacancies are the dominant defect species in BaCd$_2$P$_2$: $V_{Ba}$ and $V_{Cd}$ are the dominant acceptors; $V_P$ is the dominant donor. When ionized, the vacancies can have low formation energies for Fermi-level positions close to the band edges. Fortunately, they are all shallow defects, and can therefore be ruled out as nonradiative recombination centers. Early on in the study of lead halide perovskites, it was hypothesized that the long carrier lifetime and "defect tolerance" of MAPI$_3$ result from the shallow nature of vacancy defects.[39,42,73-75] While for MAPI$_3$ this picture has been challenged by recent studies,[76-78] BaCd$_2$P$_2$ clearly shows shallow vacancies for both cations and anions. The formation of shallow vacancies has been linked to the electronic structure of the material and, more specifically, to the presence of an antibonding upper valence band and a bonding lower conduction band.[38,42,79] Indeed, for BaCd$_2$P$_2$, most of the upper valence bands consist of antibonding states between Cd 4$d$ and P 3$p$, and the lower conduction bands show an overall bonding character which is a mix of bonding interactions between Cd 4$p$, Ba 4$d$, and Ba 6$s$ (though antibonding interactions between Cd 4$d$ and P 3$p$ are also observed) (Figure S6). We note that the bonding character of the lower conduction bands of BaCd$_2$P$_2$ arising from two types of cations is not common, suggesting that chemical bonding can be more complex in materials and that interaction beyond cation-anion orbital overlap can determine the electronic structure.

Some of the antisite defects in BaCd$_2$P$_2$ can also be low in formation energy, notably for Ba$_{Cd}$ and Cd$_{Ba}$. Given the large difference in ionic radii between Cd$^{2+}$ and Ba$^{2+}$ (1.09 Å vs 1.49 Å), one would expect that it should be energetically much less favorable to form Ba$_{Cd}$ than Cd$_{Ba}$. However, our previous analysis has shown no clear correlation between ionic size difference and antisite formation energy.[36] What our previous study found, however, is that elements with similar preferred oxidation states tend not to form antisites with deep levels. This is the case here for Ba occupying the site of smaller Cd but surprisingly not for Cd replacing the larger Ba. The Ba$_{Cd}$ introduces no defect level in the band gap of BaCd$_2$P$_2$, as expected. By contrast, Cd$_{Ba}$ is a deep double acceptor. This may be understood from the larger antisite space of Cd$_{Ba}$, in which not only Cd$^{2+}$ but also Cd$^{+}$ and Cd$^{0}$ with bigger ionic size can form. Both the P$_{Cd}$ and Cd$_{P}$ antisites are amphoteric with deep levels. The amphoteric behavior of these two antisites may be linked to the oxidation behavior of P which can act both as a cation and anion and have oxidation states spanning from +5 to −3. The interstitials have overall high formation energies and should not be a concern for nonradiative carrier recombination. Based on these results, we will later assess quantitatively the effectiveness of Cd$_{Ba}$, P$_{Cd}$, and Cd$_{P}$ as nonradiative recombination centers. We note that cation-anion antisites have also been found to form deep defects in other phosphide solar absorbers such as InP[80] and Zn$_3$P$_2$.[81]

The defect formation energies plotted in Figures 3A–3B and S5 provide important insights into doping in BaCd$_2$P$_2$. Irrespective of the elemental chemical potentials, BaCd$_2$P$_2$ clearly will not be highly doped either $p$-type or $n$-type by the intrinsic point defects. The equilibrium Fermi-level position, which can be estimated to be near the crossing point between the formation-energy curves for the dominant acceptor and donor species (such as $V_{Cd}$ and $V_P$ in Figure 3A), never gets close to the band edges, implying very low free carrier concentration. Extrinsic acceptor doping could potentially make BaCd$_2$P$_2$ $p$-type doped, especially under the P-rich condition where the $V_P$ donor and other donor defects have overall high formation energies meaning weak compensation (Figure 3B). In contrast, $n$-type doping in BaCd$_2$P$_2$ is likely to be challenging, due to strong compensation from $V_{Ba}$ and $V_{Cd}$, whose formation energies can drop to nearly or even below zero for Fermi-level positions close to the conduction-band minimum (CBM). This doping asymmetry agrees with the relatively high absolute energy of the valence-band edge of BaCd$_2$P$_2$ (see Table S1).[82,83] We conclude that BaCd$_2$P$_2$ is a natively semi-insulating material with high potential of realizing good $p$-type doping. It could be used as an intrinsic layer in a $p$-$i$-$n$ junction or as a $p$-type layer in a $p$-$n$ junction.

## Deep defects and nonradiative recombination in BaCd₂P₂

Strong nonradiative carrier recombination requires deep defects with a high enough concentration but also large carrier capture coefficients.[84-86] From the above (Figures 3A–3B), the deep defects with low formation energies are Cd$_{Ba}$, P$_{Cd}$, and Cd$_P$. The capture coefficients can be calculated from first principles considering nonradiative transition between the deep defect level and an electron at the CBM or a hole at the valence-band maximum (VBM) via multiphonon emission (details in Note S2). Here, we calculate capture coefficients for Cd$_{Ba}$, P$_{Cd}$, and Cd$_P$, and determine nonradiative carrier recombination rates in BaCd₂P₂. We assume an intrinsic BaCd₂P₂ absorber layer in a *p-i-n* device.

The Cd$_{Ba}$ gives rise to a deep (0/−) level at 0.89 eV above the valence band (Figures 3A–3B). However, the configuration coordinate diagram for Cd$_{Ba}$ (0/−) transition indicates that the electron capture barrier by $Cd_{Ba}^0$ and the hole capture barrier by $Cd_{Ba}^-$ are both large (Figure S7A). Therefore, without explicitly computing the capture coefficients, Cd$_{Ba}$ should not be a strong recombination center. The P$_{Cd}$ has a (+/−) transition level near the mid gap. The direct transition between −1 and +1 states will not be efficient, due to the low probability for a defect capturing two electrons (or two holes) at once.[87] The recombination process of P$_{Cd}$ can be treated by including the unstable neutral charge state ($q = 0$) and through two steps involving the (0/−) and (0/+) transition levels, as previously done for the iodine interstitial (I$_i$) in MAPI₃.[88] This amounts to four capture processes and coefficients in total: $C_n^0$ and $C_n^+$ for electron ($n$) capture and $C_p^0$ and $C_p^-$ for hole ($p$) capture, as shown in Figure 3C. The P$_{Cd}$ (0/−) and (0/+) levels are also very deep and locate 0.71 and 0.78 eV above the VBM, respectively (Figures S8A–S8B). Figures 3D and 3E show the configuration coordinate diagrams for the P$_{Cd}$ (0/−) and (0/+) transitions, respectively. All four capture processes have small energy barriers, which qualitatively indicates that the capture processes by P$_{Cd}$ can be easy. Quantitatively, for P$_{Cd}$, our calculations including electron-phonon coupling yield a total capture coefficient ($C_{tot}$) of $2.77 \times 10^{-7}$ cm³/s at 300 K (see Figure S8C), which is significant but comparable to that of the most active recombination center (I$_i$) in MAPI₃.[86,88] The local atomic structures of $P_{Cd}^0$, $P_{Cd}^-$, and $P_{Cd}^+$ plotted in Figure 3C highlights large displacements of the phosphorus antisite as a result of the charge transitions: the electron (hole) capture by $P_{Cd}^0$ leads to a downward (upward) displacement of the phosphorus antisite, and $P_{Cd}^-$ ($P_{Cd}^+$) shows a trigonal planar (trigonal pyramidal) P₄ configuration with three equidistant P-P bonds. The Cd$_P$ antisite gives rise to a deep (0/2+) level as well. Compared to P$_{Cd}$, Cd$_P$ presents a negligibly small $C_{tot}$ at 300 K, mainly due to large hole capture barriers (Figures S7B–S7C; Table S3).

The above results indicate that P$_{Cd}$ is the most efficient nonradiative recombination center in BaCd₂P₂, followed by Cd$_P$, and that Cd$_{Ba}$ is inactive. In the following, we determine the nonradiative recombination rates in intrinsic BaCd₂P₂ at 300 K (see Note S2). The nonradiative recombination rate (denoted as $A$) is given by $A = C_{tot} \times N_d$, where $C_{tot}$ and $N_d$ are the total capture coefficient and concentration of a recombination center, respectively. Figure 3F plots the nonradiative recombination rate $A$ in intrinsic BaCd$_2$P$_2$ as a function of the elemental chemical potentials which control defect formation energy and concentration. The recombination rate is larger under P-rich conditions which favor the formation of P$_{Cd}$. Remarkably, the recombination rate is at most $\sim 10^5$ s$^{-1}$, which are orders of magnitude smaller than that of MAPI$_3$ ($\sim 10^7$ s$^{-1}$, computed at a similar level of theory).[86,88] This is because the P$_{Cd}$ concentration is rather low (on the order of $10^{11}$ cm$^{-3}$ or less) in intrinsic BaCd$_2$P$_2$. Indeed, the lowest formation energy of P$_{Cd}$ even under the extreme P-rich condition is higher than 1.0 eV (see Figure 3A), which is still larger than the highest formation energy of the I$_i$ in MAPI$_3$ under iodine-rich condition.[88] We also emphasize that the $C_{\text{tot}}$ of P$_{Cd}$ is comparable to that of I$_i$ in MAPbI$_3$. Using the calculated $A$ rates, the nonradiative lifetime ($= 1/A$) in intrinsic BaCd$_2$P$_2$ is on the order of at least 10 $\mu$s, indicating the potential for remarkably long bulk carrier lifetime.

If BaCd$_2$P$_2$ is used as a *p*-type absorber layer in a *p-n* junction, the carrier capture behavior changes. In *p*-type BaCd$_2$P$_2$, the electron capture coefficients will dominate the total capture coefficient (Note S2; Table S3). As a consequence, not only P$_{Cd}$ but also Cd$_P$ will act as nonradiative centers. In addition, the P$_{Cd}$ and Cd$_P$ concentrations will be orders of magnitude higher when the Fermi level gets closer to the valence band (Figures 3A–3B). Assuming that the Fermi level is 0.2 eV above the VBM, the nonradiative recombination rates are $\sim 10^6$–$10^8$ s$^{-1}$ (Figure S9). Such values are still on par with those in standard inorganic *p*-type solar absorbers CdTe,[89,90] CIGS,[91] and CZTS,[92,93] where the computed $A$ rates of $\sim 10^7$ s$^{-1}$ have been typically reported.

## Synthesis and stability of BaCd$_2$P$_2$

All of our computational results (band gap, optical absorption, carrier mobilities, intrinsic point defects, and nonradiative carrier recombination at deep defects) highlight BaCd$_2$P$_2$ as a potentially high-performance inorganic solar absorber made of low-cost elements. BaCd$_2$P$_2$ was reported experimentally, but no experimental characterization of the optoelectronic properties of this material had been made.[53,54] We have synthesized BaCd$_2$P$_2$ powder samples by solid-state reaction of the elemental Ba, Cd, and P in stoichiometric ratio in sealed silica ampoules. Details about the synthesis and characterization of the crystal structure and phase stability are provided in Notes S3–S6 of the supplemental information. Figure 4A plots the powder X-ray diffraction (PXRD) pattern of the BaCd$_2$P$_2$ sample, which shows no coexistence of secondary phases and confirms the previously reported crystal structure. The crystal lattice parameters obtained from PXRD Rietveld refinement agree well with previous measurements and our calculations (see Table S4).

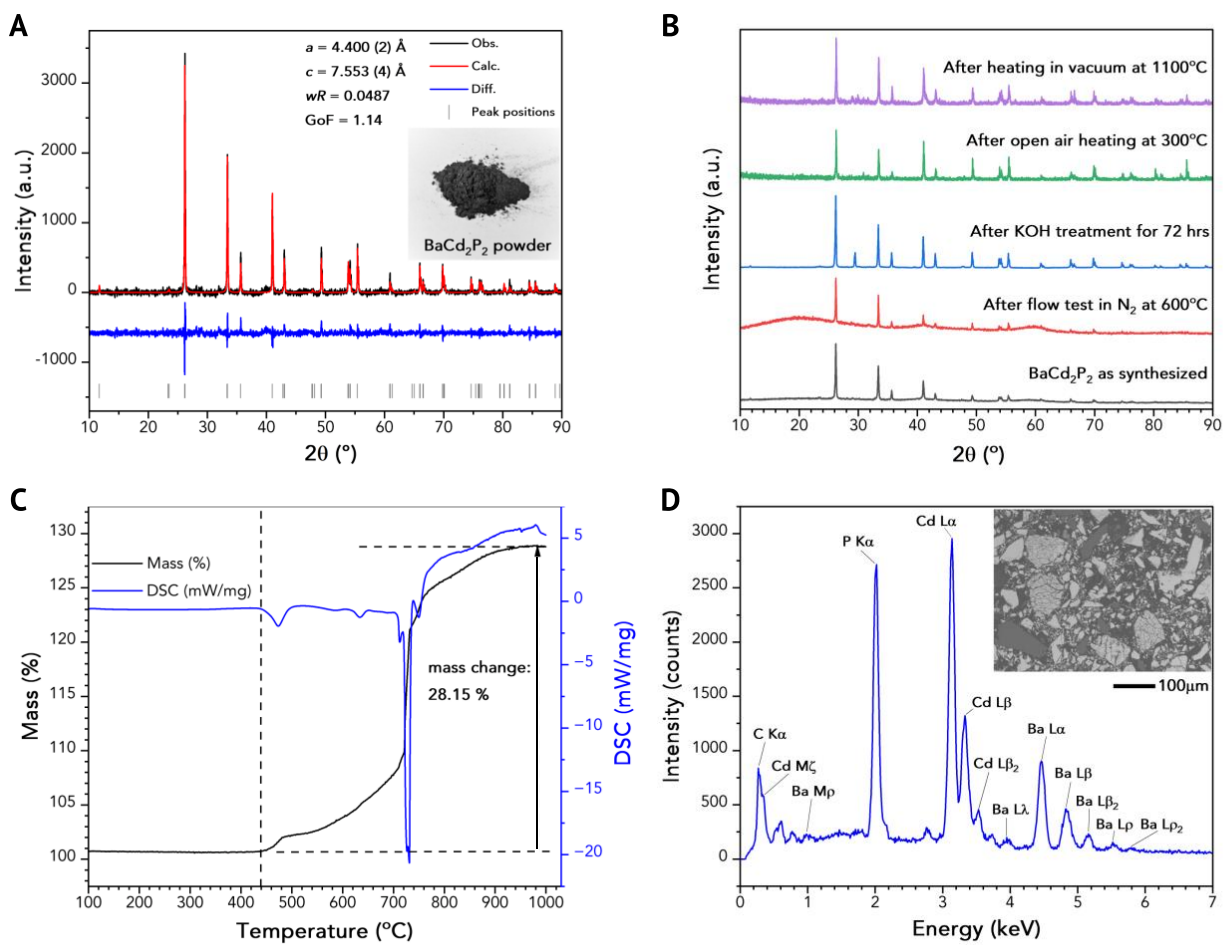

**Figure 4 Characterization of crystal structure and phase stability of BaCd$_2$P$_2$ polycrystalline sample**
(A) PXRD pattern with Rietveld refinement for as-synthesized BaCd$_2$P$_2$ sample. (B) PXRD patterns of the BaCd$_2$P$_2$ samples after stability tests under various conditions. (C) TGA and DSC analysis of the thermal behavior of BaCd$_2$P$_2$ in air atmosphere. (D) Representative EDX spectra of BaCd$_2$P$_2$ with SEM image.

As highlighted by halide perovskites whose applications are limited by their stability,[10-12] air- and moisture-sensitivity has become an important concern for the technological deployment of new solar cell absorbers. We find that BaCd$_2$P$_2$ is highly stable in ambient air (for more than 6 months at room temperature), and that immersing a sample in water for 12 hours causes no appreciable changes in the PXRD pattern. Besides, BaCd$_2$P$_2$ is also found to be resistant to 2.5M KOH solution for at least 72 hours, probably due to passivation of surface with hydroxide as indicated by the appearance of a small PXRD peak at 29° corresponding to Cd(OH)$_2$ (Figure 4B). Only strong acids such as HCl, concentrated HNO$_3$, and *aqua regia* can dissolve BaCd$_2$P$_2$. PXRD pattern of BaCd$_2$P$_2$ after heating in open air at 300 ºC shows no significant change in crystal structure (Figure 4B). PXRD taken after a flow test in N$_2$ gas reveals the sample to be stable even at 600 ºC. In vacuum, annealing at temperatures up to 1100 °C results in no changes attested by the alike PXRD patterns. During thermogravimetric analysis (TGA) and differential scanning calorimetry (DSC) in ambient atmosphere, BaCd$_2$P$_2$ does not show any appreciable change until 425 ºC after which the material starts to oxidize (Figures 4C and S10–S11). Morphology and overall stoichiometry of the synthesized BaCd$_2$P$_2$ are examined using scanning electron microscopy (SEM) and electron-dispersive X-ray spectroscopy (EDS) (Figures 4D and S12–S13). The EDS analysis reveals a nearly stoichiometric BaCd$_2$P$_2$ sample and importantly no impurity phases in the sample. Overall, all these results demonstrate a very stable BaCd$_2$P$_2$ material, for which oxidation is probably kinetically limited and which should not require complex encapsulation and cumbersome processing.

## Photoluminescence (PL) and carrier lifetime in BaCd$_2$P$_2$ powder

We have performed optoelectronic characterization of the BaCd$_2$P$_2$ powder samples. PL spectra are collected using a micro-PL setup in air, which allows us to target individual grains and conduct a statistical analysis of the PL intensity (see details in Note S7). Figure 5A shows a pronounced PL peak at 1.46 eV (847 nm) at 298 K, in very good agreement with the HSE band gap (1.45 eV) as well as the step in the transmittance spectra of BaCd$_2$P$_2$ (shown in the same figure). This suggests that the PL peak at 847 nm is due to band-to-band radiative recombination. A weaker PL peak at 980 nm (1.27 eV) is also observed, probably due to a shallow defect level close to the band edge. Characterizing this second peak will be the focus of our future work. Notably, the intensity of the PL peaks only slightly decreases when the sample is heated up to 368 K, and is fully recovered upon cooling back to 298 K (Figure 5A). The robust and strong PL spectra confirms that BaCd$_2$P$_2$ is also highly stable in terms of its optoelectronic properties in the entire temperature range of 298–323 K for typical solar cell operation.[94]

To assess the PL efficiency more quantitatively and benchmark it versus a well-studied solar cell material, we compare in Figure 5B the PL spectra of BaCd$_2$P$_2$ and GaAs powders under identical excitation conditions. The GaAs sample is obtained by pulverizing a piece of device-grade single-crystal prime GaAs (001) wafer into a fine powder. GaAs is chosen because it is a well-established high-efficiency solar material and has a similar band gap. The BaCd$_2$P$_2$ and GaAs powder samples have similar particle sizes (Figure S14), so we expect comparable degrees of surface recombination and multiple scattering optical losses in both samples. Remarkably, Figure 5B shows that the PL peak intensity of our unoptimized BaCd$_2$P$_2$ is already in the same order as that of high-purity prime GaAs (i.e., only ∼3 times weaker). This is also clearly seen in Figure 5B inset which shows the statistical histograms for 100 random spots on each sample. The result is encouraging, since BaCd$_2$P$_2$, as a new solar material, lacks the level of purity of the prime single-crystal GaAs wafer.

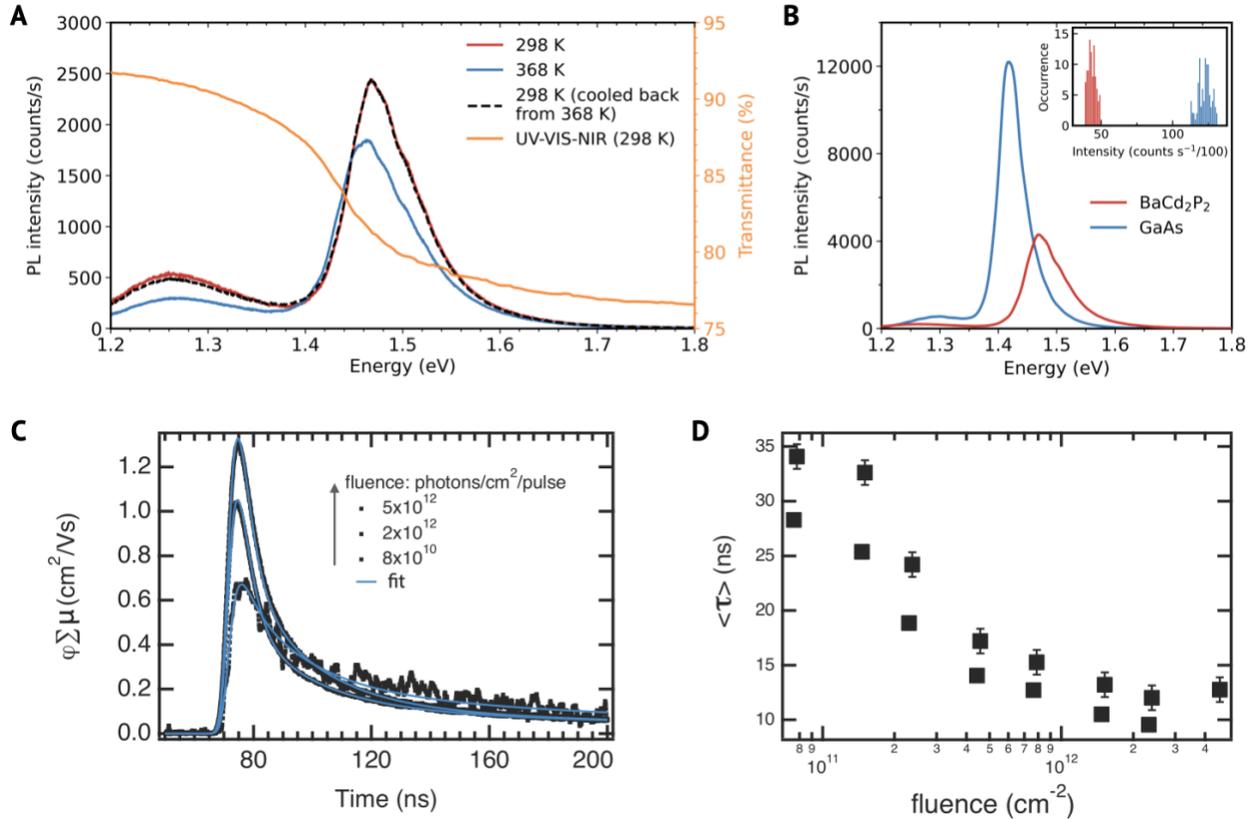

**Figure 5 Photoluminescence (PL) and microwave conductivity measurements of BaCd$_2$P$_2$ polycrystalline samples** (A) Comparison between PL spectra at 298 K, 368 K, and after cooling down the sample from 368 K back to 298 K. (B) Average micro-PL spectra of BaCd$_2$P$_2$ powder versus that of GaAs powder. The inset histograms count the occurrence of the PL intensity across the powders (C) Three representative photoconductivity transients (black) and tri-exponential fits (blue) as a function of laser fluence (600 nm, 7 ns FWHM), expressed as the product of carrier yield ($\varphi$) and mobility sum ($\Sigma \mu$). (D) Population-weighted average carrier lifetime $\langle \tau \rangle$ extracted from the tri-exponential fit to all transients collected for two sample replicates as a function of incident laser fluence.

For a more quantitative study of carrier transport and excited-state dynamics in BaCd$_2$P$_2$, we have performed time-resolved and steady-state microwave conductivity (TRMC[95,96] and SSMC[97]) measurements on BaCd$_2$P$_2$ powder (see Note S8). Microwave conductivity measurements employ a microwave field to probe the conductive and dielectric properties of semiconductor materials without the need for electrical contacts, making it particularly suitable for characterizing powder samples of emerging materials before high-quality thin films are available. Here, we analyze the equilibrium (i.e., dark) properties of the BaCd$_2$P$_2$ powder (Figure S15A), its transient carrier dynamics (Figures 5C–5D and S15B), and its photoconductivity action spectrum (Figure S16). Figure 5C shows photoconductivity transients expressed as a product of charge carrier yield and mobility sum for three representative laser fluences (see also Figure S15B). The blue solid lines show a fit to the data employing three exponential decay components, from which the population-weighted average carrier lifetime is calculated. As seen in Figure 5D, the carrier lifetime is within 10–30 ns depending on fluence; lower fluences result in longer lifetimes. An up to 30 ns photoconductivity lifetime is promising for unoptimized BaCd$_2$P$_2$ powders, since the surface recombination effects are likely at play. For comparison, such a carrier lifetime already exceeds those measured for CdTe (before 2014)[98] and CZTS (present) thin-films.[99] Even halide perovskites, which are well known for their long carrier lifetimes, started with lifetimes of a few to a few hundred ns measured on thin films.[100,101] As discussed above, carrier lifetime is strongly correlated with the efficiency potential and prospect of a new solar cell absorber.[102]

The yield-mobility product extracted from TRMC provides a weighted sum of the electron and hole mobilities,[103,104] assuming that every absorbed photon creates an unbound electron-hole pair ($\varphi = 1$). Here, we calculate its value using the sum of the pre-exponential factors from the fits in Figure 5C so as to account for recombination that occurs within the instrument response function.[105] These values are also fluence-dependent. Interestingly, the yield-mobility product *increases* with fluence, from 1 cm$^2$/Vs at the lowest fluence to 4 cm$^2$/Vs at the highest fluence (see also Figure S15B). Similar behavior is observed in steady-state photo-modulation (Figure S16B), where photoconductivity rises superlinearly ($\Delta\sigma \propto F^{1.5}$) with light flux ($F$). Under most circumstances the steady-state photoconductivity is expected to be either linear (first-order recombination) or sublinear ($\Delta\sigma \propto F^{0.5}$ for bimolecular recombination). Both our observations suggest that the present mobility and lifetime results are limited by a significant population of trap sites: as the injection density increases and traps become filled, the average carrier mobility increases.[106] This observation could be associated with the high-surface-area of the powder under investigation (Figure 4D inset). The steady-state photoconductivity action spectra (Figure S16A) and the PL spectra (Figure 5A) are consistent with this surface trap hypothesis. The former exhibits a non-zero tail to the red of the evident absorption onset near 855 nm (i.e., 1.45 eV), and the latter shows a broad second emission peak in the same region.

Finally, we note that the powder measured by TRMC has a XRD coherence length of only ~100 nm (Note S8), which likely suggests that the microwave frequency-dependent mobilities reported above are severely limited by the scattering effects due to crystalline grain boundaries or extended structural defects. Indeed, using a confinement length of 100 nm and an observed 9.6-GHz mobility of 4 cm$^2$/Vs we get an intrinsic mobility of ~100 cm$^2$/Vs (Figure S17), in agreement with the computed mobilities. It is worth noting that in the early-stage development of halide perovskite solar cells, the carrier mobilities in MAPI$_3$ were also within the order of magnitude measured for BaCd$_2$P$_2$ but increased with improvements in film quality.[107]

## Conclusions

We have computationally screened about 40,000 materials searching for high-efficiency, stable, and low-cost thin-film solar cell absorbers. A handful of solar absorber candidates have emerged and among them BaCd$_2$P$_2$ stands out for its computed direct band gap (1.45 eV), high carrier mobilities, large optical absorption coefficients, and most importantly favorable point-defects which will not lead to strong nonradiative carrier recombination. BaCd$_2$P$_2$ is also made of low-cost and non-critical elements. Carrier capture calculations show that P$_{Cd}$ is the dominant nonradiative recombination center, but its high formation energy leads to a very long nonradiative lifetime, at least two orders of magnitude higher than that of MAPI$_3$.

Our synthesized BaCd$_2$P$_2$ powder samples show an experimental band gap in good agreement with theory and exhibit bright PL on par with the luminescence of a high-purity GaAs wafer powder. TRMC measurements on unoptimized powder indicate a carrier lifetime of up to 30 ns, which surpasses early halide perovskites thin films and is on par with present-day established inorganic PV absorbers. This experimentally confirms the promises of BaCd$_2$P$_2$ as a new thin-film solar absorber. In addition to its exceptional optoelectronic and defect properties, BaCd$_2$P$_2$ shows extremely high stability in air and water. It is stable in air for more than 6 months at room temperature, and oxidation only starts for temperature above 400 °C. The high stability will benefit the future development of BaCd$_2$P$_2$ solar cells.

The attractive properties of BaCd$_2$P$_2$ as a thin-film absorber material motivate the future growth of high-quality thin films and fabrication of solar cells with either a *p-n* or *p-i-n* junction device architecture. We note that several other phosphides crystalize in the BaCd$_2$P$_2$ structure type including the less toxic Zn-based CaZn$_2$P$_2$ and SrZn$_2$P$_2$. In view of the exceptional properties exhibited by BaCd$_2$P$_2$, these zinc-based compounds are of interest for future studies. The large family of Zintl BaCd$_2$P$_2$ structure type materials can offer opportunities for tuning band gap and other properties through elemental substitutions.

Beyond BaCd$_2$P$_2$, our work highlights how high-throughput computational screening including point-defect properties and follow-up experimental synthesis and characterizations can identify unexpected solar absorbers with exceptional intrinsic PV properties among tens of thousands of materials. It paves the way for more computation-driven materials discovery in the PV field.

## Acknowledgments

This work was supported by the U.S. Department of Energy, Office of Science, Basic Energy Sciences, Division of Materials Science and Engineering, Physical Behavior of Materials program under award number DE-SC0023509 to Dartmouth and was authored in part by the National Renewable Energy Laboratory, operated by Alliance for Sustainable Energy, LLC, for the U.S. Department of Energy (DOE) under Contract No. DE-AC36-08GO28308. All computations, syntheses, and characterizations were supported by this award unless specifically stated otherwise. This research used resources of the National Energy Research Scientific Computing Center (NERSC), a DOE Office of Science User Facility supported by the Office of Science of the U.S.

Department of Energy under contract no. DE-AC02-05CH11231 using NERSC award BES-ERCAP0023830. A. P. acknowledges support from a Department of Education GAANN fellowship. Funding for microwave conductivity measurements and analysis on semiconductor powders provided by the Materials Chemistry Program, Materials Sciences and Engineering Division, Office of Basic Energy Sciences, U.S. Department of Energy under grant DE-SC0023316. PPMS instrument used for initial attempts of conductivity measurements was supported by Ames National Laboratory, U.S. Department of Energy, which operates under the Contract DE-AC02-07CH11358. The views expressed in the article do not necessarily represent the views of the DOE or the U.S. Government.